\documentstyle[twocolumn,aps,epsf]{revtex}
\begin{document}
\twocolumn[
\begin{center}
{\large{\bf Vortex states of the ${\mbox{\boldmath $E_{u}$}}$ model 
for Sr${\mbox{\boldmath $_{2}$}}$RuO${\mbox{\boldmath $_{4}$}}$}}

\vspace{5mm}
{Takafumi Kita}
\par

{\small\em Division of Physics, Hokkaido University, 
Sapporo 060-0810, Japan}
\par

({\today})
\vspace{3mm}

{\small \parbox{142mm}{\hspace*{5mm}
Based on the Ginzburg-Landau functional of $E_{u}$ symmetry presented by
Agterberg, vortex states of Sr$_{2}$RuO$_{4}$ are studied in detail over $%
H_{c1}\!\lesssim \!H \!\leq\! H_{c2}$ by 
using the Landau-level expansion method. For the
field in the basal plane, it is found that (i) the second superconducting
transition should be present irrespective of the field direction; (ii) below
this transition, a characteristic double-peak structure may develop in the
magnetic-field distribution; (iii) a third transition may occur between two
different vortex states. It is also found that, when the field is along the $%
c$ axis, the square vortex lattice may deform through a second-order
transition into a rectangular one as the field is lowered from $H_{c2}$.
These predictions will be helpful in establishing the $E_{u}$ model for Sr$%
_{2}$RuO$_{4}$.
}
}
\end{center}

\vspace{8mm}
]

Active studies have been performed on superconducting
Sr$_{2}$RuO$_{4}$\cite{Maeno94}
where another unconventional pairing may be realized\cite{Rice95}.
A possible candidate for its symmetry is the $E_{u}$ model
with two-fold degeneracy\cite{Agterberg97},
as indicated by various experiments
\cite{Mackenzie98,Ishida97,Ishida98,Jin99,Honerkamp98,Yamashiro98,Luke98}. 
However, further experiments seem being required
before establishing its validity for Sr$_{2}$RuO$_{4}$.
In this respect, the vortex states may provide clear and indisputable tests for
the {\em p}-wave hypothesis.
The present paper provides a detailed
theoretical description of them which will be helpful towards
that purpose.
Clarifying the basic features of the two-component model,
which has not been performed completely,
will also be useful for the experiments of UPt$_{3}$.

The vortex states of the $E_{u}$ model for Sr$_{2}$RuO$_{4}$ have been
studied theoretically in a series of papers by Agterberg {\em et al}.\cite
{Agterberg98-1,Agterberg98-2,Heeb99}. Based on the two-component
Ginzburg-Landau (GL) functional and following essentially Abrikosov's method 
\cite{Abrikosov57} which is effective near the upper ($%
H_{c2}$) and lower ($H_{c1}$) critical fields, they have provided several
important predictions. Especially noteworthy among them are: existence of
the second transition for ${\bf H}\!\perp\!{\bf c}$ similar to
that observed in UPt$_{3}$\cite{Bruls90,Adenwalla90}; several
orbital-dependent phenomena helpful in identifying which band is mainly
relevant; stabilization of the square vortex lattice for ${\bf H}%
\!\parallel\!{\bf c}$. An observation of the square lattice
has been reported by Riseman {\em et al}.\cite{Riseman98}.

With these results, this paper focuses on the following: (i) The
properties of the intermediate fields, in particular those below the
second transition for ${\bf H}\!\perp\!{\bf c}$, remain to be
clarified. We will treat the whole range $H_{c1}\!\leq\! H\!\leq\! H_{c2}$
in a unified
way, describe possible changes of experimentally detectable properties as a
function of the field strength, and draw characteristic features in low
fields. (ii) Considered have been the cases where the field is along
the high-symmetry axes. It is still not clear whether or not the second 
transition
for ${\bf H}\!\perp\!{\bf c}$ persists for arbitrary field
directions in the $ab$ plane, because the term
$|\eta _{1}|^{2}|\eta _{2}|^{2}$
in the GL functional [see Eq.\ (\ref{freeEn}) below] generally causes the 
first- and third-order mixing. We will study those general cases to establish the
existence of the second transition. 
(iii) Agterberg introduced several assumptions in the parameters used
to minimize the free energy. We will perform the minimization without 
such assumptions.

The goals (i)-(iii) may seem rather formidable, but they can be achieved
with the Landau-level expansion method\cite{Kita98-1}. When applied to the $s
$-wave pairing, it successfully reproduced the properties of the whole
region $\ H_{c1}\!\lesssim\! H\!\leq\! H_{c2}$ quite efficiently for an arbitrary $%
\kappa $. Compared with the direct minimization procedure in real space\cite
{Machida93}, the method has a couple of advantages that (i) it is far more
efficient and (ii) one can enumerate possible second-order transitions\
rather easily, hence enabling us to establish the phase diagram of various
multi-order-parameter systems.
This is the first time where it is applied to 
an multi-order-parameter system so that this paper also has
some methodological importance.

The GL free-energy density adopted by Agterberg is given by\cite
{Agterberg98-1} 
\begin{eqnarray}
f=\! &&-|\mbox{\boldmath $\eta$}|^{2}+\frac{1}{2}|\mbox{\boldmath $\eta$}%
|^{4}+\frac{\gamma }{2}(\mbox{\boldmath $\eta$}\times \mbox{\boldmath$\eta$}%
^{\ast })^{2} +(3\gamma \! -\! 1)\,|\eta _{1}|^{2}|\eta _{2}|^{2}\vspace{3mm} 
\nonumber \\
&&-\, \mbox{\boldmath $\eta$}^{\dagger }\left[ 
\begin{array}{cc}
D_{x}^{2}\!+\!\gamma D_{y}^{2}\!+\!\kappa _{5}D_{z}^{2} & \gamma
(D_{x}D_{y}\!+\!D_{y}D_{x})\vspace{3mm} \\ 
\gamma (D_{x}D_{y}\!+\!D_{y}D_{x}) & D_{y}^{2}\!+\!\gamma
D_{x}^{2}\!+\!\kappa _{5}D_{z}^{2}
\end{array}
\right] \mbox{\boldmath $\eta$}  \nonumber \\
&&+\, h^{2}\, ,  \label{freeEn}
\end{eqnarray}
where the same notations are used here. This
simplified free energy
has the advantage that there are only two parameters in it whose values can
be extracted from experiments, i.e.\ $\kappa _{1}\!\equiv \!H_{c2}/\sqrt{2}%
H_{c}$ and $\nu \!\equiv \!\frac{1-3\gamma }{ 1+\gamma }$, the latter 
being related
to the $H_{c2}$ anisotropy in the $ab$ plane as $H_{c2}({\bf a}%
)/H_{c2}({\bf a}+{\bf b})\!=\!\frac{1-\nu }{1+\nu }$\cite{Agterberg98-1}. The
value $\kappa _{1}\!=\!31$ $(1.2)$ for ${\bf H}\!\perp\!{\bf c}$ ($%
{\bf H}\!\parallel\!{\bf c}$) will be used throughout\cite
{Yoshida96}, whereas $\nu $ is left as a parameter. A recent observation of
the $H_{c2}$ anisotropy suggests that $\nu $ is positive 
and $\sim 0.01$\cite{Mao99}.

We sketch the method to find the minimum for an arbitrary field strength\cite
{Kita98-1}. Let us fix the mean flux density ${\bf B}\equiv (B\sin \theta
\cos \varphi ,B\sin \theta \sin \varphi ,B\cos \theta )$ rather than the
external field $H$, and express ${\bf h}\!=\!{\bf B}+\widetilde{{\bf h}}$ where the
spatial average of $\widetilde{{\bf h}}$ vanishes by definition. We then
transform
\begin{equation}
\left[ 
\begin{array}{c}
x \\ 
y \\ 
z
\end{array}
\right] \!= \!\left[ 
\begin{array}{ccc}
\cos \theta \cos \varphi & -\sin \varphi & \sin \theta \cos \varphi \\ 
\cos \theta \sin \varphi & \cos \varphi & \sin \theta \sin \varphi \\ 
-\sin \theta & 0 & \cos \theta
\end{array}
\right] \!\!\left[ 
\begin{array}{c}
x^{\prime }/L \\ 
y^{\prime }L \\ 
z^{\prime }
\end{array}
\right] ,  \label{trans1}
\end{equation}
\begin{equation}
\mbox{\boldmath $\eta$}({\bf r)}=\! \left[ 
\begin{array}{cc}
\cos \frac{\phi }{2} & -\sin \frac{\phi }{2} \\ 
\sin \frac{\phi }{2} & \cos \frac{\phi }{2}
\end{array}
\right]\! \left[ \!
\begin{array}{cc}
\cos \phi' & \sin \phi'  \\ 
-i\sin \phi'  & i\cos \phi' 
\end{array}
\! \right] \mbox{\boldmath $\eta$}^{\prime }({\bf r}^{\prime }{\bf )} \, ,
\label{trans2}
\end{equation}
where $\phi$, $\phi'$ and $L$
are conveniently chosen as 
$\phi\! =\!\tan ^{-1}\left[ {2\gamma }\tan 2\varphi /({1-\gamma })
\right] $, $\phi' =L^2 \cos\theta$, 
and $L=\left\{ (1\!+\!\gamma \!-\!f)/[(1\!+\!\gamma \!-\!f)\cos ^{2}\theta
+2\kappa _{5}\sin ^{2}\theta ]\right\} ^{1/4}$ with $f\!\equiv \!\left[
(1\!-\!\gamma )^{2}\cos ^{2}2\varphi +4\gamma ^{2}\sin
^{2}2\varphi \right] ^{1/2}$.
Assuming uniformity along ${\bf z}^{\prime }$ direction,
we then expand  $\mbox{\boldmath
$\eta$}^{\prime }({\bf r}^{\prime }{\bf )}$ and $\widetilde{{\bf h}}$ $({\bf %
r}^{\prime }{\bf )}$ as 
\begin{equation}
\mbox{\boldmath $\eta$}^{\prime }({\bf r}^{\prime }{\bf )=}\sqrt{V}\sum_{N%
{\bf q}}{\bf c}_{N{\bf q}}\, \psi _{N{\bf q}}({\bf r}^{\prime }) \, ,
\label{expand1}
\end{equation}
\begin{equation}
\widetilde{{\bf h}}({\bf r}^{\prime })=\widehat{{\bf z}}^{\prime }\sum_{{\bf %
K\neq 0}}\widetilde{h}_{{\bf K}}\, \exp(i{\bf K}\cdot{\bf r}^{\prime})
\, ,  \label{expand2}
\end{equation}
where $V$ is the system volume, $\psi _{N{\bf q}}$ denotes an
eigenstate of the magnetic translation group in the flux density $B$ with
the Landau-level index $N$ and the magnetic Bloch vector ${\bf q}$, and $%
{\bf K}$ is the reciprocal lattice vector of the vortex lattice. The
explicit expression of $\psi _{N{\bf q}}({\bf r}^{\prime })$ for the spacial
case where one of the unit vectors of the vortex lattice, ${\bf a}_{2}$,
lies along the $y^{\prime }$ axis is given by 
\begin{eqnarray*}
&& \psi _{N{\bf q}}({\bf r}^{\prime })\!=\!\!\!
\sum_{n=-{\cal N}_{{\rm f}}/2+1}^{{\cal N}_{%
{\rm f}}/2}\!\!\!\!{\rm e}^{i[q_{y^{\prime }}(y^{\prime }\!+0.5q_{x^{\prime
}})+na_{1x^{\prime }}(y^{\prime }\!+q_{x^{\prime
}}-0.5na_{1y^{\prime }})]/l_{c}^{2}}  \nonumber \\
&&\times \sqrt{\frac{2\pi l_{c}/a_{2}}{2^{N}N!\sqrt{\pi }\,V }}H_{N}\!\!\left(\! 
\frac{x^{\prime }\!-\! q_{y^{\prime }}\! -\! na_{1x^{\prime }}}{l_{c}}\!\right) 
{\rm e}^{-(x^{\prime }\!- q_{y^{\prime }} 
- na_{1x^{\prime }})^{2}/2l_{c}^{2}}
\end{eqnarray*}
with ${\cal N}_{{\rm f}}^{2}$ the number of flux quanta in the system, $l_{c}$
denoting $\frac{1}{\sqrt{2}}$ of the magnetic length, and $a_{1x^{\prime }}$
($a_{1y^{\prime }}$) the $x^{\prime }$ ($y^{\prime }$) component of another
unit vector ${\bf a}_{1}$\cite{Kita98-1}. We
also consider the counterclockwise rotation of 
${\bf a}_{1}$ and ${\bf a}_{2}$
around the $z^{\prime }$ axis by the angle $\varphi _{L}^{\prime }$.
Substituting Eqs.\ (%
\ref{trans1})-(\ref{expand2}) into Eq.\ (\ref{freeEn}) and integrating over the
volume, we obtain the free energy per unit volume as 
\begin{equation}
F[\left\{ {\bf c}_{N{\bf q}}\right\} ,\{\widetilde{h}_{{\bf K}}\},B,\rho
,\vartheta ,\varphi _{L}^{\prime }]=\frac{1}{V}\!\!\int \!\!f[%
\mbox{\boldmath
$\eta$}^{\prime }({\bf r}^{\prime }),\widetilde{{\bf h}}({\bf r}^{\prime
})\!,B]d^{3}r^{\prime }  \label{freeEn2}
\end{equation}
where $\rho \!\equiv \! |{\bf a}_{1}|/|{\bf a}_{2}|$ and $\vartheta \!\equiv
\!\cos ^{-1}\!\frac{{\bf a}_{1}\cdot {\bf a}_{2}}{|{\bf a}_{1}||{\bf a}_{2}|}
$. This $F$ is a desired functional which can be minimized rather easily
using one of the standard minimization algorithms\cite{NumRec}. Due to the
periodicity of the vortex lattice, we only have to perform the integration
over a unit cell. The external field $H$ is then determined through the
thermodynamic relation ($H\!=\!\frac{1}{2}\frac{\partial F}{\partial B}$ in
the present units). In numerical calculations we have cut the series in
Eqs.\ (\ref{expand1}) and (\ref{expand2}) at some $N_{c}$ and $|{\bf K}_{c}|$%
, respectively, thereby obtaining a variational estimate of the free energy.
The convergence can be checked by increasing $N_{c}$ and $|{\bf K}_{c}|$.
The choice $N_{c}\sim 12$ and ${\bf K}_{c}\sim $(the third smallest) has
been checked to provide correct identification of the free-energy minimum
with the relative accuracy of $10^{-6}$ for $B/H_{c2}\!\gtrsim\! 0.1$. 
Though not
presented here, preliminary calculations reveal that the method is also
effective for $\theta\!\neq\! 0$, $\frac{\pi}{2}$.

The functional $F$ has another advantage that one may enumerate possible
transitions in the vortex states of multi-order-parameter systems.
Much attention has been focused on this subject in connection with
the observed phase diagram of superconducting UPt$_{3}$\cite{UPt-theo}. No
complete analysis has appeared yet, however, and the use of $F$ will be
quite helpful for that purpose. 
The features of the
$s$-wave lattice 
can be summarized as follows\cite{Kita98-1}: (a) a
single ${\bf q}$ in Eq.\ (\ref{expand1}) suffices to describe it with a
choice of ${\bf q}$ corresponding to the broken translational symmetry of
the lattice; (b) the hexagonal (square) lattice is made up of $N\!=\!
6n$ ($4n$) Landau levels ($n$: integer)\cite{Ryan,Kita98-1}; (c) more
general structures can be described with $N\!=\! 2n$ levels, odd $N$'s
never mixing up since those bases have finite amplitude at the core sites;
(d) the expansion coefficients ${\bf c}_{N{\bf q}}$ can be chosen real
for the hexagonal and square lattices. With these results on the
conventional lattice, the following second-order transitions are possible in
multi-component systems: (i) deformation of the hexagonal or square
lattice which accompanies entry of new $N$'s as well as complex numbers in
the expansion coefficients; (ii) mixing of another wave number ${\bf q}_{2}$
satisfying ${\bf q}_{2}\!-\!{\bf q}_{1}\!=\!{\bf K/}2$\cite{UPt-theo,Garg94};
(iii) entry of odd $N$'s. Though not complete,
this consideration will be sufficient below.

\begin{figure}[t]
\begin{center}
\leavevmode
\epsfxsize=70mm
\epsfbox{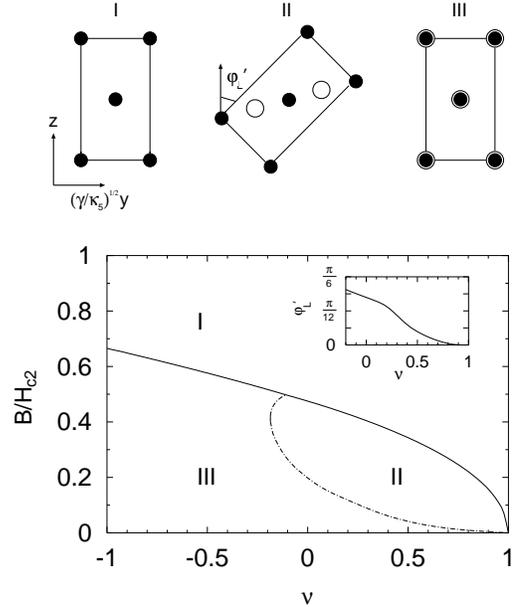}
\end{center}
\caption{Transition lines for ${\bf H}\!\parallel\!{\bf a}$ as a function
of the anisotropy parameter $\nu$. The closed (open) circles denote the zeros 
of $\eta_1$ ($\eta_2$). The inset plots 
the angle $\varphi_{L}^{\prime}$ at the I$\leftrightarrow$II transition
as a function of $\nu$.}
\label{fig:1}
\end{figure}\newpage

\begin{figure}[t]
\begin{center}
\leavevmode
\epsfxsize=35mm
\epsfbox{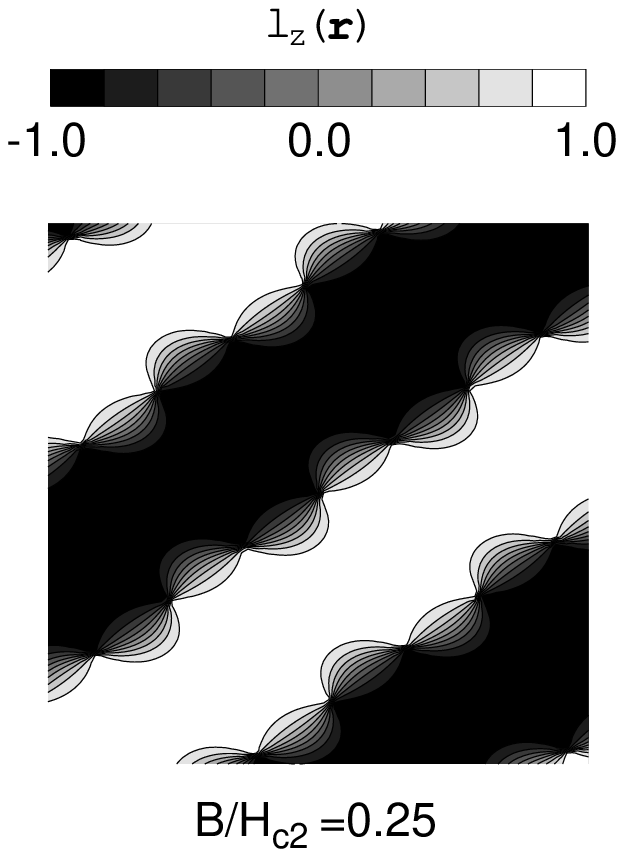}
\end{center}
\caption{Spatial variation of $l_{z}\!\equiv \!
(\mbox{\boldmath $\eta$}\!\times \!
\mbox{\boldmath$\eta$}^{\ast})\cdot \widehat{{\bf z}}/2i|\eta_1||\eta_2|$
in the state II for $\nu\!=\! 0.077$ and $B/H_{c2}\!=\! 0.25$.
The regions with $l_{z}\!\approx \!\pm 1$ correspond to the ``bulk'' state.
See Fig.\ \ref{fig:3} for the corresponding  $|%
\mbox{\boldmath $\eta$}({\bf r})|$.}
\label{fig:2}
\end{figure}
\begin{figure}[t]
\begin{center}
\leavevmode
\epsfxsize=80mm
\epsfbox{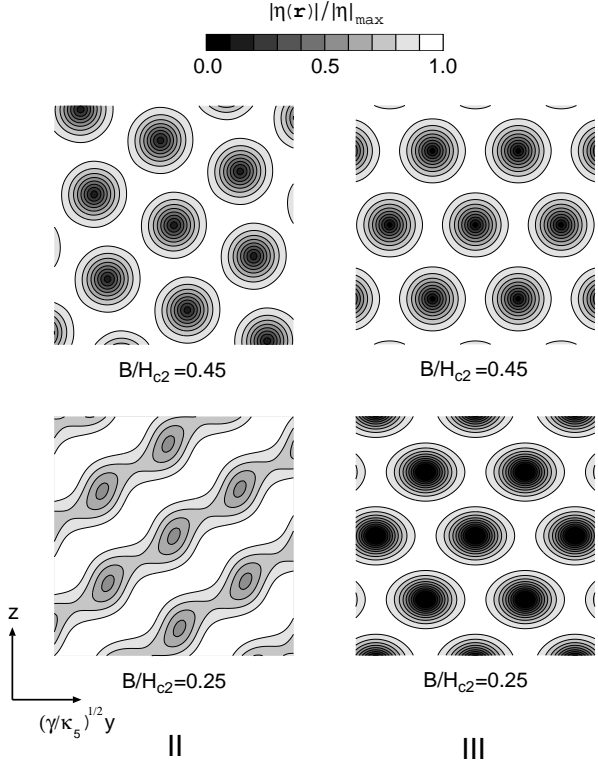}
\end{center}
\caption{A comparison of $|%
\mbox{\boldmath $\eta$}({\bf r})|$ between the states II and III
with $\nu\!=\! 0.077$ and $B/H_{c2}\!=\! 0.45$, $0.25$.
The amplitude $|\mbox{\boldmath $\eta$}({\bf r})|$ is finite 
everywhere in the state II, 
which is brought about at the expense of the variation in $l_{z}$; 
see Fig.\ \ref{fig:2}.}
\label{fig:3}
\end{figure}

We now present the results for ${\bf H}\!\perp\!{\bf c}$. Figure 1
shows the transition lines for ${\bf H}\!\parallel\!{\bf a}$ ($%
\theta\!=\!\frac{\pi}{2}$; $\varphi\!=\! 0$)
as a function of the anisotropy parameter $\nu$;
the one given as a function of $\gamma =\frac{1-\nu }{3+\nu }$ has
qualitatively the same structure, with $\gamma\! =\! 0$  and  
$1$  respectively 
corresponding  to  $\nu\!=\! 1$  and  $-1$. 
As already pointed out by Agterberg
\cite{Agterberg98-1}, there are three possible vortex states: the high-field
region I where a hex-
\begin{figure}[t]
\begin{center}
\leavevmode
\epsfxsize=80mm
\epsfbox{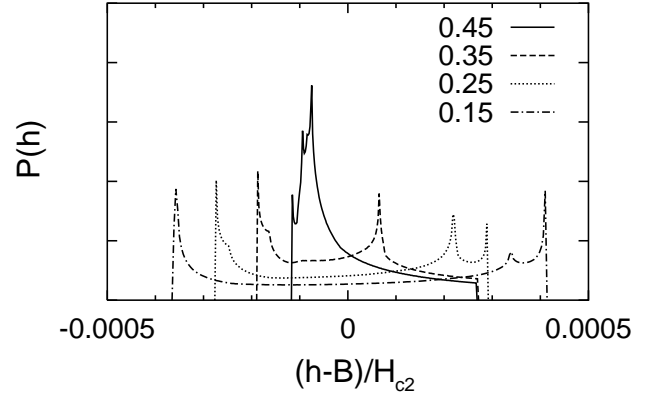}
\end{center}
\caption{The magnetic-field distribution $P(h)$ in the state II
for $B/H_{c2}\!=\! 0.45$, $0.35$, $0.25$, and $0.15$.
A characteristic double-peak structure develops as the field is decreased.}
\label{fig:4}
\end{figure}
\noindent
agonal lattice is stable with $\eta _{2}\! =\! 0$; the region II where 
$\eta _{2}$ becomes finite with ${\bf q}%
_{2}\!-\!{\bf q}_{1}\!$ equal to half the unit vector ${\bf b}_{1}$ of the
reciprocal lattice, i.e.\ the vortex lattice is coreless with $|%
\mbox{\boldmath$\eta$}|$ finite everywhere; the region III where a deformed
conventional lattice with $\eta _{2}\!\neq\! 0$ is stable.
In addition, Fig.\ 1 includes the following new results: 
(i) a full minimization with respect to $\varphi _{L}^{\prime}$
clarifies that the I$\leftrightarrow $II transition is continuous as a
function of $\nu $  (see the inset); (ii) high-precision
calculations in the low-field region reveal that, as the field is lowered,
the coreless state II is replaced via a first-order transition by the state
III with cores. The reason for (ii) can be realized by looking at the
variation of $l_{z}\equiv (\mbox{\boldmath $\eta$}\!\times \!
\mbox{\boldmath$\eta$}^{\ast})\cdot \widehat{{\bf z}}/2i|\eta_1||\eta_2|$ 
which is proportional to the magnitude of the orbital
angular momentum along ${\bf z}$. As seen in Fig.\ 2
calculated for $\nu\!=\! 0.077$ ($\gamma\!=\! 0.3$) and 
$B/H_{c2}\!=\! 0.25$,
one of the bulk states $l_{z}\!=\!\pm\! 1$ is alternately realized
in II, and 
there necessarily exist lines of ``defects'' where $l_{z}$
vanishes. Compared with III where $|\mbox{\boldmath $\eta$}|$
vanishes at points, the state II is thus energetically unfavorable at low
fields. It can however be stabilized at intermediate fields by making $|%
\mbox{\boldmath $\eta$}|$ \ more uniform. Figure 3 plots $|%
\mbox{\boldmath $\eta$}({\bf r})|$ \ for $\nu =0.077$,
showing how the differences between II and III develop as $B/H_{c2}$ is
decreased. In fact, only a deformation of the lattice occurs
in III,
whereas a layered structure also shows up in II with $|%
\mbox{\boldmath $\eta$}({\bf r})|$ becoming more and
more uniform.
This rather drastic change in II can be detected
by measuring the magnetic-field distribution $P(h)\equiv \frac{1}{V}\!
\int \delta [h-h_{x}({\bf r})]\,d^{3}r$. As seen in
Fig.\ 4, the single peak at $B/H_{c2}\!=\! 045$ 
splits and one of them moves
towards the high-field end, which originates from the development
of a ridge in $h_{x}({\bf r})$ along a valley of $|%
\mbox{\boldmath $\eta$}({\bf r})|$. The observation of it
by NMR or $\mu $SR experiments will form
a direct evidence for the state II as
well as for the presence of multi-order parameters.
It is also quite interesting to perform the experiments in UPt$_{3}$ 
where a lattice distortion has already been detected\cite{Yaron97}. 
We finally point out that the second-order transition between I$\leftrightarrow
$II or I$\leftrightarrow $III is present for an arbitrary field direction in
the basal plane.\hfill A glance on the functional (\ref{freeEn}) may

\begin{figure}
\begin{center}
\leavevmode
\epsfxsize=75mm
\epsfbox{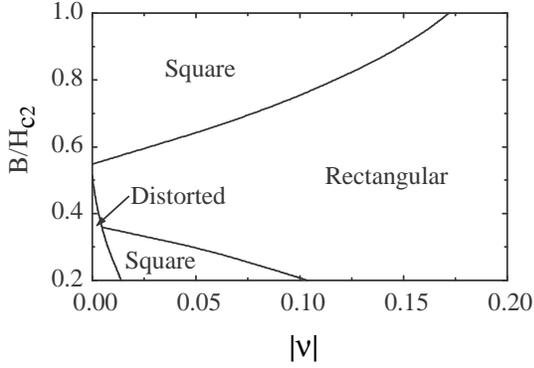}
\end{center}
\caption{The vortex-lattice phase diagram for ${\bf H}\!\parallel\!{\bf c}$ 
as a function of $|\nu|$ and $B/H_{c2}$ for $\kappa_1 \! =\! 1.2$. The angle $\varphi_{L}'$ is 
$\frac{\pi}{4}$ and $0$ for $\nu\!>\! 0$ and $\nu<\! 0$, 
respectively.}
\label{fig:5}
\end{figure}
\noindent
 lead to the
conclusion that the transition I$\leftrightarrow $III disappears for a
low-symmetry direction, since the term $|\eta_{1}|^{2}|\eta _{2}|^{2}$
yields those like $\eta _{1}^{\prime }\eta _{2}^{\prime \ast }|\eta
_{1}^{\prime }|^{2}$. However, it does persist as the transition (i)
classified in the preceding paragraph.The hexagonal lattice has been checked to be
stable in the high-field region, and the phase diagram for a small $|\nu |$
is qualitatively similar to Fig.\ 1.

We finally present the results for ${\bf H}\!\parallel\!{\bf c}$.
Figure 5 shows the vortex lattice structure as 
a  function  of 
$|\nu|$  and  $B$ 
for $\kappa _{1}\!=\!1.2$. The
square lattice is stabilized near $H_{c2}$ for small values of $|\nu |$,
confirming Agterberg's result through a perturbation expansion with respect
to $\nu$ ($\kappa _{1}\!=\!1.2$ corresponds Agterberg's $\kappa \!\sim\!
0.66$ for $\nu\!=\!0$)\cite{Agterberg98-2}. 
As the field is decreased, however,
the lattice deforms into a
rectangular one for $|\nu |\lesssim 0.17$,
followed by a further transition into the square and/or a distorted 
(i.e.\ $\rho\!\neq\! 1$; $\vartheta\!\neq\!\frac{\pi}{3},\frac{\pi}{2}$)
lattice for $|\nu|\lesssim 0.1$. The same calculation
for $\kappa _{1}\!=\! 2.6$ reveals that all the phase boundaries move 
rightward, with the distorted, square, and rectangular regions extending 
over $0\!\leq\! B/H_{c2} \!\leq\! 1$ for $|\nu|\!\lesssim\! 0.004$,
$0.02\!\lesssim\!|\nu|\!\lesssim \! 0.09$, and $ 0.23\!\lesssim\! |\nu|$,
respectively.
With $\kappa _{1}$ and $|\nu |$ small,
the free energies of these
lattices are not much different from one another, as suggested by
Agterberg's $\nu $-$\kappa $ diagram near $H_{c2}$\cite{Agterberg98-2}, and
the present calculation reveals that there may also be field-dependent
transformations among them. Although Riseman {\em et al}.\cite{Riseman98}
have reported an observation of the square lattice, there may exist
field-dependent distortion in the diffraction pattern. A detailed
experiment on the field dependence may be worth carrying out.

The author is grateful to M.\ Sigrist for several useful conversations,
and to D.\ F.\ Agterberg for valuable comments on the original manuscript.
Numerical calculations were performed on an Origin 2000 in "Hierarchical
matter analyzing system" at the Division of Physics, Graduate School of
Science, Hokkaido University.

\end{document}